# A Theoretical Framework for Simulating Organizations

*Edmundo Barrientos Palma, edmundo.barrientos@usach.cl*
*Universidad de Santiago de Chile*
*Departamento de Ingeniería Industrial*
*Av. Ecuador 3769, Santiago de Chile*

**Abstract**

This work proposes a theoretical framework using a systemic modeling paradigm to implement computational agents in the simulation of organizations. The potential of its use is demonstrated in the modeling of supply chains. Finally, research tending to develop an organizational modeling system in real-time is proposed.

## Introduction

It is not a mystery that currently, human activity organizations, specifically those linked to generating profits, are subjected to daily demands that they must face to remain competitive (Porter, 1985).

Thus, intra and extra-organizational knowledge is necessary to allow managers, at any level and function, to improve decision-making to achieve the company's global objectives.

For forecasting future situations, in the scientific world, especially the engineering world, a commonly used tool is modeling "reality" (Simon, 1990). The quality of the decisions chosen depends to a large extent on this model.

One approximation is the use of tools generated thanks to research in the field of Computational Organization (CO) for the modeling and subsequent simulation of organizations based on Multiagent Systems (Weiss and Sen, 1999), (Multi-Agent Systems, MAS).

The proposal presented in this paper uses a systemic modeling paradigm, which will be compared with the current paradigm, the reductionist analytic, for the representation of communities of agents, applied later in the simulation of supply chains.

Finally, it is proposed to expand the concept of simulation for the development of real-time modeling systems, that is, the obtaining and updating of data at different levels in information systems implemented from the model, such as product demands, production parameters, sales, etc.; allowing a link with proposals in the field of Management Control and ERP (Enterprise Resource Planning) systems.

## Computer Organization

In research in the field of CO, computational and mathematical methods are used to theorize and analyze organizations and their organizational process in their human and automatic components (Carley and Prietula, 1994).

The objective of the CO is to build theories about organizations, and develop tools and methodologies for the validation and analysis of organizational computational models, reflecting these models in practice and obtaining a better understanding of the organization. Organizations are heterogeneous, complex, and composed of systems or agents (human or artificial), who make decisions, and carry out activities depending on their capacities and rules. Are dynamic, non-linear, and adaptive, which, when interacting, result in the organization. This interaction is described through structures and networks: authority or formal structure, friendship network, task structure, and skill structure, for example. (Carley and Prietula, 1994).

It is based on theories such as Distributed Artificial Intelligence (DAI), multi-agent systems (MAS),

adaptive agents, organizational theory, communication theory, social networks, and information diffusion.

These computational models can be used to demonstrate the legitimacy of various accepted theories in Organization Theory, testing multiple configurations or scenarios (simulations) to make sense of the organization's objectives.

There are various levels of detail, and therefore of tools, in modeling organizations. The simplest is used to test general principles of organizational behavior, and the most specific to examine the performance of certain organizations in the face of changes, such as re-engineering and acquisition of new technologies.

There are various implementations of multi-agent systems (Weiss and Sen, 1999), such as MACE, SDML, SOAR, and SWARM.

Some applications of the Computational Organization:

- Carley and Tsvetovat (2002) test the hypothesis that markets tend to self-segment into sub-markets.

- Chang and Harrington (2000), where a computational model of a retail chain is presented to determine how the amount of discretion given to warehouse managers, as well as how they run their businesses, influences innovation, concluding that greater decentralization improves the performance of the firm, when the storage markets are sufficiently different, the horizon is long enough and the markets are stable.

Traditionally, the methods used in model validation are based on comparing its outputs with reality, with other models, with model results in degenerate situations, etc. However, when what is being modeled is a complex system, such as an organization of human activity, these methodologies are not widely accepted (Brown and Kulasiri, 1996).

## Systemic Modeling

The proposal is based on Jean-Louis Le Moigne's General Theory of Systems (Le Moigne's Systemics or LMS), studied by Eriksson (1997). It is based on three questions about knowledge and therefore modeling, the "what," the "why," and the "how":

- What is knowledge? is based on two hypotheses: the phenomenological and the teleological.

    - Phenomenological: tells us that knowledge is the action of the "knower," who builds artificial representations of the interaction between the object and the subject. Thus, knowledge does not exist because it is or because things are "so" instead, it emerges from those interactions. In addition, the cognitive interactions between the experience of the subject and the subject form the knowledge organized by itself. The phenomenological proposition has 3 properties:

        - Dialectic: representing the interaction between subject and object. Operationally it implies the use of the analytical or atomistic approach in addition to the synthetic or holistic approach.

        - Irreversibility: knowledge is an action rather than a state; it has a temporary property.

        - Recursion: indicates the self-reference quality of knowledge, allowing the "knower" to accept the cognitive act of self-reference, forgotten from the Aristotelian logic of excluding third parties.

• Teleological: tells us that the constitution of knowledge is based on the desires, objectives, and aspirations of the knowledge generators; that is, they depend on the class identity or viability of the subjects.

• Why is knowledge valid? Presents the criteria to validate knowledge, projective or collective viability, contrasted with the positivist aspiration of objective truth.

• How to obtain knowledge?, where cognitive tools for the constitution of knowledge are presented, including the Nine Levels Model (NML)-

Le Moigne (1990) proposes the NLM, Figure 1, as a reference to conceptualize systems:

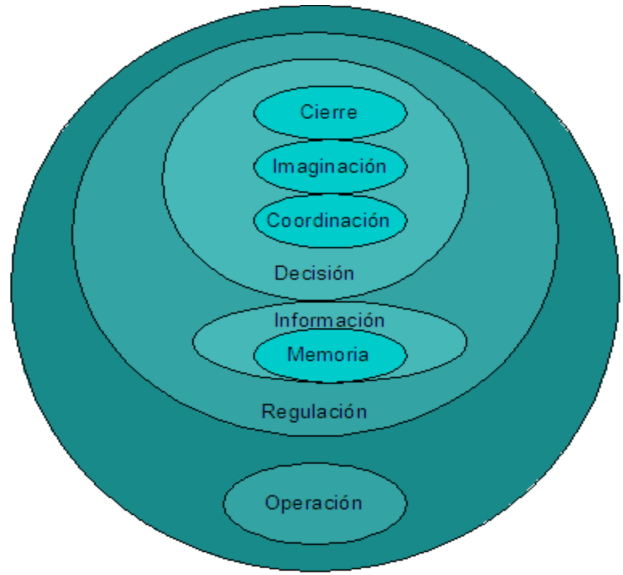

*Figure 1: Nine Levels Model*

Level 1: "In an environment, the observer distinguishes a system."

Level 2: "The observer distinguishes what the system does."

Level 3: "The observer postulates the existence of regulation mechanisms."

Level 4: "The observer postulates the existence of information flows for regulation."

Level 5: "The observer postulates the existence of a decision system for their behaviors."

Level 6: "The observer postulates the existence of a system that memorizes the information."

Level 7: "The observer postulates the existence of a system that coordinates his action decisions."

Level 8: "The observer postulates the existence of a system that imagines and conceives new possible decisions."

Level 9: "The observer postulates the existence of a system that grants closure and completion."

## Proposal: Theoretical Framework to Implement Multiagent Systems

The proposal in this work uses software agents implemented under an approach based on the systems modeled through the NLM to simulate organizational phenomena, mainly.

The Java language is used to implement the primary interfaces, referencing the different levels of the NLM. Notwithstanding the use of other programming languages (c++) or tools (Arena), object-oriented can be used.

The class diagram is as follows:

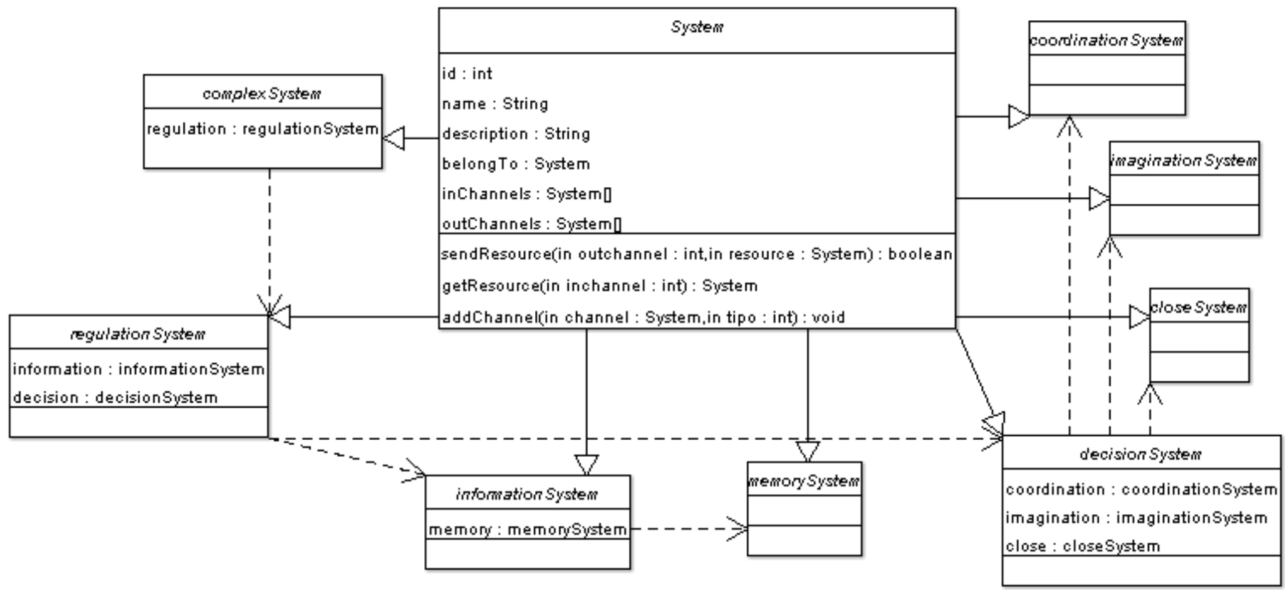

Each of the systems (implemented by the "System" class) must be observed as a black box (Ashby, 1997), with input channels that will allow the established output (agent function, level 2 NLM) through channels. Channels are systems (list "inChannel" and "outChannel" inside the "System" class). These input and output flows can be information, material, or energy, represented by agents ("System" class). Thus, we can define agents without their regulation subsystem (level 3 NLM), that is, only the implementation of the operational subsystem (level 2 of the NLM). According to Tarride (2003), all systems, when recognized, are regulated because they maintain the essential variables within the margins allowed to be such; without the proposal being in opposition to said statement, if not instead the project modeling concept is applied, that is, the systems must be modeled to validate them, up to a level that allows their projective feasibility, not being important the recognition of the level of regulation.

The "doing" of the agents must be made explicit in the operational subsystem (level 2 of the NLM and "System" class) by defining functions that receive and send other agents, representing information, material, or energy. In the "regulationSystem" class, we must define the essential variables of the modeled phenomenon.

The information subsystem, belonging to the regulation system, receives the information flows from the system, that is, what is captured by its "sensors," defined according to the utility for the viability of the system or agent, that is, in the utility for its "doing." Sensors are systems that capture information outside the system, so in simulation terms, they are the mechanisms to detect disturbances as inputs to the operations system ("System" class). These disturbances can be stored in the memory subsystem ("memorySystem" class, level 6 of the NLM).

In the decision subsystem, level 5 of the NLM, and class "decisionSystem," the classes "coordinationSystem," "imaginationSystem," and "closeSystem" must be implemented. In "coordinationSystem", level 7 of the NLM, coordination algorithms must be implemented in decision-making, which may include assigning priorities, etc. The "imaginationSystem" class, representing level 8 of the NLM, must consist of algorithms that allow the search for patterns in the information subsystem, especially in historical details, that is, the memory subsystem.

Finally, at level 9 of the NLM, or the "closeSystem" class, completions must be implemented in the decisions. These can be of the if-then type.

There are support libraries for the construction of multi-agent systems. In this work, we use Repast (Tobias, 2003), which allows the visualization of flexible parameters belonging to the agents, regardless of their implementation type. It also includes a kit of predefined functions for managing the simulation iterations (pause, stop, run) and the graphic display of variables belonging to the different agents.

The simulation model must be adaptable over time, that is, be adjustable to meet new demands, this being permissible thanks to the teleological approach to its construction, which also implies the interaction of various actors in the modeled phenomenon who have different visions about it, arguing for the improvement of the model (projective validity).

## Supply Chain Modeling

According to van der Zee and van der Vorst (2005), the needs for modeling the supply chain as a complex system are:

- Modeling of elements and their relationships: assigning control policies to members and specific relationships within the supply chain, such as hierarchy and coordination, imply explicit attention to decision variables.

- Modeling of the dynamics: the time and the execution of the decision activities must be explicit, for example, in the stock levels and dispatch times.

- User interface: the active participation of those in charge of the supply chain is necessary to impregnate veracity in the solution among all the parties involved and to improve the quality of this solution.

- Ease of scenario planning: given the supply chain's complexity, many possible scenarios may be required, where reusable models can help increase the speed of modeling and subsequent analysis.

The current approaches for the simulation of supply chains focus on the representation of "physical interactions," relegating to implicit and dispersed modeling in the model, the entities related to coordination and control; Furthermore, this is valid in the context of manufacturing (van der Zee and van der Vorst, 2005).

Van der Zee and van der Vorst (2005) propose a theoretical framework for supply chain simulation using the multi-agent systems paradigm.

The proposal presented in this paper meets the needs stated above. Still, unlike the van der Zee and van der Vorst proposal, it aims at simulating organizational systems, whatever they may be: manufacturing, supply chain, organizational structure. , etc. Therefore, its main advantage lies in integrating the various organizational processes at multiple levels.

## Conclusions

This proposal for modeling organizational processes can be complemented, with a temporary vision of it, that is, visualizing the future (simulation) with experiences and data, past and present. Level 6 of the NLM makes explicit reference to the memory of a system, so it is possible to store information in database systems (represented as agents), obtained from "sensors" (level 4 of the NLM), distributed within and

Outside of an organization, they can be artificial or human, allowing the construction and visualization in real-time of essential variables or management indicators of the subsystems, constituting the necessary knowledge for their survival (Piaget, 1969).

The tools built according to this approach would allow the creation of information systems from the modeling of the organization, being a technical challenge, the automatic construction of the applications necessary for the exploitation of said systems, for which the use of the base code could be released under open licenses (GNU, Creative Commons), in a modular design that allows the construction of "systems" that can be coupled (Ashby, 1997), improved and shared.

The modules can belong to a low level (concerning the deployment of complexity) or operational, that model specific machinery containing the necessary interface (software and hardware) to transmit the information to the distributed system, up to a high level, where complete systems are modeled, such as a supply chain, relegating the problems of ERP systems, in terms of their generality, or integrating these software tools into the distributed system.

Traceability can be added to the products using non-regulated agents and Radio Frequency Identification (RFID) technologies, which will send the information of the products to the distributed system.

Finally, it is indicated that this proposal's empirical testing and validation is necessary.